# Nonlinear Autoregressive Approach to Estimating Logistic Model Parameters of Urban Fractal Dimension Curves

Yanguang Chen

(Department of Geography, College of Urban and Environmental Sciences, Peking University, 100871, Beijing, China. Email: chenyg@pku.edu.cn)

**Abstract:** A time series of fractal dimension values of urban form can form a fractal dimension curve and reflects urban growth. In many cases, the fractal dimension curves of cities can be modeled with logistic function, which in turn can be used to make prediction analysis and stage division studies of urban evolution. Although there is more than one method available, it is difficult for many scholars to estimate the capacity parameter value in a logistic model. This paper shows a nonlinear autoregressive approach to estimating parameter values of logistic growth model of fractal dimension curves. The process is as follows. First, differentiating logistic function in theory with respect to time yields a growth rate equation of fractal dimension. Second, discretizing the growth rate equation yields a nonlinear autoregressive equation of fractal dimension. Third, applying the least square calculation to the nonlinear autoregressive equation yields partial parameter values of the logistic model. Fourth, substituting the preliminarily estimated results into the logistic models and changing it into a linear form, we can estimate the other parameter values by linear regression analysis. Finally, a practical logistic model of fractal dimension curves is obtained. The approach is applied the Baltimore's and Shenzhen's fractal dimension curves to demonstrate how to make use of it. This study provides a simple and effective method for estimating logistic model parameters, and it can be extended to the logistic models in other fields.

# 1 Introduction

Urban growth and form is one of the important subjects in urban studies. Urban growth is linked geographical process, and urban form is linked to geographical pattern and spatial structure. The two form a pair of conjugate concepts. However, urban form has no characteristic scale and cannot be effectively described by using the conventional measure such as perimeter, area, and density. Fractal geometry provides a powerful tools for characterizing scale-free geographical phenomena such as cities (Batty and Longley, 1994; Frankhauser, 1994). A great number of studies show that urban form can be well described with fractal dimension. A series of fractal dimension in different times composes a sample path of time series. If the sample path is long enough, we can use it to generate a curve reflecting urban growth. Due to squashing effect of fractal dimension increase, a fractal dimension curve of urban growth can be modeled by sigmoid function (Chen, 2012; Chen, 2018). Among the family of various sigmoid functions, the simplest and the most common one is logistic function. The indispensable step of logistic modeling of fractal dimension curves for researching fractal city development is parameter estimation.

It is easy to calculate the parameter values of the common two-parameter logistic model. However, it is inconvenient to estimate the parameter values of the three-parameter logistic model. For the two-parameter logistic model, the value of the capacity parameter is known. Therefore, a two-parameter logistic model can be readily converted into a simple linear relationship, and thus the univariate linear regression analysis can be employed to estimate model parameter values. For the three-parameter logistic model of fractal dimension curves of urban form, the key method of parameter estimation rests with working out the value of the capacity parameter. The method of curve fitting of nonlinear function can be utilized to estimate the parameters of the general logistic model. But there are two problems. On the one hand, not all researchers can write the computer program of curve fitting; on the other hand, this method receives the more significant influence of the numbers with larger size in a dataset, so that the modeling results are not accurate enough to describe the early stage of urban development. In this paper, a nonlinear autoregressive method is advanced to estimate the parameter value of the logistic model of fractal dimension curves of urban form. The other parts are organized as follows. In Section 2, the necessary mathematical equations and formulae are derived for the parameter estimation. In Section 3, three sample paths of fractal dimension of two cities, Baltimore and Shenzhen, are utilized as examples to show how to make use of the approaches. In Section 4, the related questions are discussed, and finally, in Section 5, the discussion is concluded with summarization of main points of this study.

## 2 Models

### 2.1 Nonlinear regressive equations and related formulae for random sampling

In the simplest case, the fractal dimension curves of urban growth can be described using the ordinary logistic function. The logistic model of fractal dimension curves of urban growth can be obtained through the idea of squashing effect (Chen, 2012). Generally, the model of fractal dimension increase can be expressed as

$$D(t) = \frac{D_{\max}}{1 + (D_{\max} / D_{(0)} - 1)e^{-kt}} ,$$  (1)

where $D(t)$ denotes the fractal dimension of urban form at time $t$, $D_{(0)}$ refers to the initial value of fractal dimension at time $t$=0, $D_{\max}$ is the capacity value of fractal dimension, i.e., the upper limit of fractal dimension, and $k$ is the initial growth rate of fractal dimension. Equation (1) can be changed into the following form

$$\ln(\frac{D_{\max}}{D(t)} - 1) = \ln(\frac{D_{\max}}{D_{(0)}} - 1) - kt = \ln a - kt .$$  (2)

Let

$$y_t = \frac{D_{\max}}{D(t)} - 1, \quad a = \frac{D_{\max}}{D_{(0)}} - 1 .$$  (3)

Thus, equation (2) can be expressed as a linear equation as follows

$$\ln y_t = \ln a - kt .$$  (4)

By means of linear regression analysis, we can estimate the values of $a$ and $k$, where $a$ is the estimated value of $D_{\max}/D_{(0)}$-1. Thus, according to equation (3), the initial value of fractal dimension as a parameter can be estimated by the formula as below

$$\hat{D}_{(0)} = \frac{D_{\max}}{a + 1} .$$  (5)

which differ slightly from the observed value of the fractal dimension at time $t$=0. In above formula, the symbol "^" indicates that the value is the result of estimation. Clearly, if the capacity value of fractal dimension, $D_{max}$, can be determined, then a simple linear regressive analysis can be made to estimate the value of $a$ and $k$. The key to solving the problem above-mentioned is to determine the value of the capacity parameter, $D_{max}$.

A generalized nonlinear autoregressive method can be employed to estimate capacity parameter value. First of all, let's consider the general situation, that is, the fractal dimension calculation time is irregular. Differentiating fractal dimension $D(t)$ in equation (1) with respect to time $t$ yields a growth rate equation

$$\frac{\mathrm{d}D(t)}{\mathrm{d}t} = kD(t)(1 - \frac{D(t)}{D_{max}}) ,$$ (6)

which represents the velocity of increase of fractal dimension of urban form. Discretizing equation (6) yields a difference equation as below

$$\frac{\Delta D(t)}{\Delta t} = kD(t)(1 - \frac{D(t)}{D_{max}}) = kD(t) - \frac{kD(t)^2}{D_{max}} = kD(t) - cD(t)^2 ,$$ (7)

in which the parameter is $c = k/D_{max}$. So we have a formula as follows

$$\hat{D}_{max} = \frac{k}{c} ,$$ (8)

which suggests that the capacity value, $D_{max}$, can be estimated by $k$ and $c$ values. In this way, two parameters can be assigned values, one is the initial growth rate, $k$, and the other is the capacity parameter, $D_{max}$.

Then, we have two methods to estimate the values of the remaining parameter(s). One is formula averaging method, and the other, is regression analysis method. The first method is to make use of averaging method by formulae. Since the values of the initial growth rate and the capacity parameter have been estimated, we only need to estimate the theoretical initial value of fractal dimension. As indicated above, the theoretical initial value differs from the realistic initial value, and the latter is known. Substituting equation (8) into equation (1) yields the initial values

$$\hat{D}_{(0)}(t_i) = \frac{\hat{D}_{max}}{1 + (\frac{\hat{D}_{max}}{D(t_i)} - 1)e^{\hat{k}t_i}} ,$$ (9)

in which $i=1,2,3,\ldots,n$ denotes the number of sampling time. The initial value of fractal dimension can be estimated by arithmetic mean, that is

$$\hat{D}_{(0)} = \frac{1}{n}\sum_{i=1}^{n}\hat{D}_{(0)}(t_i) = \frac{1}{n}\sum_{i=1}^{n}\frac{\hat{D}_{max}}{1+(\frac{\hat{D}_{max}}{D(t)}-1)e^{\hat{k}t_i}} . \qquad (10)$$

At this stage, we have obtained the estimated values of all parameters in the logistic model.

The second method is to make use of linear regression analysis to estimate the values of parameters $k$ and $D_{(0)}$. Although the $k$ value has been worked out through equation (7), it can be given up. Substituting the estimated value of the capacity parameter based on equation (7) into equation (2), we can make a linear regression analysis. By the linear regression, we can estimate the values of $a$ and $k$, where $a$ is the estimated value of $D_{max}/D_{(0)}$-1. In practice, the $k$ value given by equation (2) differs slightly from the result from equation (7). Thus the initial value of fractal dimension can be estimated by equation (5). Please note that, although in theory, the same parameter should take the same numerical value, in practice, the estimated value of $D_{(0)}$ based on equation (5) differs to a degree from that based on equation (10). Substituting the parameter values into equation (1), we can conduct city growth prediction analysis or stage division analysis.

## 2.2 Nonlinear regressive equations and related formulae for regular sampling

If the fractal dimension sampling points are regularly distributed, for example, the fractal dimension value is estimated once a year, or every other year, or every two years, we will have $\Delta t$=constant, and $\Delta D(t)= D(t+\Delta t)-D(t)$. Thus, equation (7) can be converted into the following expression

$$D(t + \Delta t) - D(t) = k\Delta t D(t) - \frac{k\Delta t}{D_{max}}D(t)^2 . \qquad (11)$$

So, the logistic model parameters can be estimated with the following equation

$$D(t + \Delta t) = (1 + k\Delta t)D(t) - \frac{k\Delta t}{D_{max}}D(t)^2 = bD(t) - cD(t)^2 , \qquad (12)$$

which is actually a nonlinear autoregressive equation. In equation (12), the regressive coefficients are $b$=1+$k\Delta t$ and $c$=$k\Delta t/D_{max}$, from which we can estimate the values of $k$ and $D_{max}$. The formulae are as below:

$$\hat{k} = \frac{b-1}{\Delta t} , \qquad (13)$$

$$\hat{D}_{\max} = \frac{\hat{k}\Delta t}{c} = \frac{b-1}{c}. \tag{14}$$

Then, we have two approaches to estimating the values of the remaining parameter(s): formula averaging method and regression analysis method. Base on equations (9) and (10), the formula averaging method can be used to estimate the value of parameter, $D_0$. Base on equations (2), (3), (4), and (5), the regression analysis method can be utilized to estimate the values of parameter $D_{(0)}$ and $k$, where the $k$ value belongs to the second estimation.

## 2.3 Nonlinear regressive equations and related formulae for continuous sampling

If the fractal dimension sampling points are continuously distributed, that is to say, fractal dimension values were calculated year by year, we will have $\Delta t = 1$, and $\Delta D(t) = D(t+1) - D(t)$. Note that the so-called continuity here is not a mathematical continuity, but an empirical continuity. The continuity of mathematics leads to differentiation rather than difference. Thus, equation (11) can be converted into the following expression

$$D(t+1) - D(t) = kD(t) - \frac{k}{D_{\max}}D(t)^2. \tag{15}$$

In this case, the logistic model parameters can be estimated by regressive analysis based on the following equation

$$D(t+1) = (1+k)D(t) - \frac{k}{D_{\max}}D(t)^2 = bD(t) - cD(t)^2, \tag{16}$$

where the regressive coefficients are $b=1+k$ and $c=k/D_{\max}$. Equation (16) is a clear nonlinear autoregressive equation. It suggests that equation (7) is essentially a nonlinear autoregressive equation. By using equation (16), we can estimate the values of $k$ and $D_{\max}$. The formulae of estimating the values of $k$ and $D_{\max}$ are as follows

$$\hat{k} = b - 1, \tag{17}$$

$$\hat{D}_{\max} = \frac{\hat{k}}{c} = \frac{b-1}{c}, \tag{18}$$

which are special cases of equations (13) and (14). By comparison, we can see the relationships, similarities, and differences between equation (8), equation (14), and equation (18).

Then, we have two methods to estimate the values of the remaining parameter(s). Similar to the previous two situations, one is the formula averaging method, and the other is the regression analysis method. The formula averaging method is as follows. The initial value of fractal dimension can be estimated by

$$\hat{D}_{(0)}(t) = \frac{D_{max}}{1 + (\frac{D_{max}}{D(t)} - 1)e^{kt}} \ .$$

(19)

The formula of the average value is

$$\hat{D}_{(0)} = \frac{1}{n}\sum_{t=1}^{n}\hat{D}_{(0)}(t) = \frac{1}{n}\sum_{t=1}^{n}\frac{\hat{D}_{max}}{1 + (\frac{\hat{D}_{max}}{D(t)} - 1)e^{\hat{k}t}} \ .$$

(20)

As for the regression analysis method, substituting the value given by equation (18) into equation (2), we can make a regression analysis to estimate the values of $D_{(0)}$ and $k$, where the $k$ value is the result from the second estimation.

**Table 1 Parameter estimation methods of logistic model of fractal dimension curve of urban growth based on combination of different equations**

| Sampling | Approach | Equation combination |
|---|---|---|
| **Random sampling** | Approach 1 | Equations (7) and (8); Equations (9) and (10) |
| | Approach 2 | Equations (7) and (8); Equations (2), (3), (4), and (5) |
| **Regular sampling,** | Approach 1 | Equations (12), (13), and (14); Equations (9) and (10) |
| | Approach 2 | Equations (12), (13), and (14); Equations (2), (3), (4), and (5) |
| **Continuous sampling** | Approach 1 | Equations (16), (17), and (18); Equations (19) and (20) |
| | Approach 2 | Equations (16), (17), and (18); Equations (2), (3), (4), and (5) |

**Note**: In approach 1 for continuous sampling sequence, equations (19) and (20) can be replaced by equations (9) and (10), equivalently.

## 2.4 Method summarization of parameter estimation

After deriving a mathematical model, corresponding parameter estimation methods can be designed. Based on the equations and formulae derived above, three sets of approaches can be designed to estimate the parameter values of logistic models of fractal dimension growth (Table 1). The first set of methods is for random sampling sequence, the second set of methods is for regular sampling sequence, and the third set of methods is for continuous sampling sequence. The method for random sequences can be used for both regular and continuous sequences, while the method for regular sequences can be used for continuous sequences. But the opposite is not true.

1. Two approaches for model parameter estimation based on random sampling sequence of fractal dimension. (1) Nonlinear bivariate auto-regression + formulae. The first approach is to make use of equation (7), (8), (9), and (10). By means of equations (7) and (8), a nonlinear auto-regression analysis can be made to estimate the values of capacity parameters $D_{max}$ and inherent growth rate $k$. Then, by using equations (9) and (10), the initial value of fractal dimension, $D_{(0)}$, can be estimated. Substituting the estimated results into equation (1) yields a logistic model of fractal dimension curve based on random sampling. (2) Nonlinear bivariate auto-regression + linear regression + formulae. The second approach is to make use of equations (2), (3), (4), (5), (7), and (8). Using equations (7) and (8) to estimate the capacity parameter $D_{max}$, and estimate the initial growth rate $k$ preliminarily. Then, using equations (2), (3), (4) and (5) to estimate the parameter $D_{(0)}$, and estimate the initial growth rate $k$ once again. Substituting the estimated results into equation (1) yields the second logistic model of fractal dimension curve based on random sampling.

2. Two approaches for model parameter estimation based on regular sampling sequence of fractal dimension. (1) Nonlinear bivariate auto-regression + formulae. The first approach is to make use of equations (9), (10), (12), (13), and (14). By means of equations (12), (13), and (14), a nonlinear auto-regression analysis can be made to estimate the values of capacity parameters $D_{max}$ and inherent growth rate $k$. Then, by using equations (9) and (10), the initial value of fractal dimension, $D_{(0)}$, can be estimated. Substituting the estimated results into equation (1) yields a logistic model of fractal dimension curve based on regular sampling. (2) Nonlinear bivariate auto-regression + linear regression + formulae. The second approach is to make use of equations (2), (3), (4), (5), (12), (13), and (14). Using equations (12), (13), and (14) to estimate the capacity parameter $D_{max}$, and estimate the initial growth rate $k$ preliminarily. Then, using equations (2), (3), (4) and (5) to estimate the parameter $D_{(0)}$, and estimate the initial growth rate $k$ once again. Substituting the estimated results into equation (1) yields the second logistic model of fractal dimension curve based on regular sampling.

3. Two approaches for model parameter estimation based on continuous sampling sequence of fractal dimension. (1) Nonlinear bivariate auto-regression + formulae. The first approach is to make use of equations (16), (17), (18), (19), and (20). By means of equations (16), (17), and (18), a nonlinear auto-regression analysis can be made to estimate the values of capacity parameters $D_{max}$ and inherent growth rate $k$. Then, by using equations (19) and (20), the initial value of fractal dimension, $D_{(0)}$, can be estimated. Equations (19) and (20) are the special cases of equations (9) and (10). So the former can be replaced with the latter. Substituting the estimated results into equation (1) yields a logistic model of fractal dimension curve based on continuous sampling. (2) Nonlinear bivariate auto-regression + linear regression + formulae. The second approach is to make use of equations (2), (3), (4), (5), (16), (17), and (18). Using equations (16), (17), and (18) to estimate the capacity parameter $D_{max}$, and estimate the initial growth rate $k$ preliminarily. Then, using equations (2), (3), (4) and (5) to estimate the parameter $D_{(0)}$, and estimate the initial growth rate $k$ once again. Substituting the estimated results into equation (1) yields the second logistic model of fractal dimension curve based on continuous sampling.

The above approaches are theoretically equivalent to one another, but empirically they may result in different parameter values. However, there is no significant difference in the same parameter's values based on different methods. All approaches involve regression and auto-regression, with nonlinear auto-regression conducted through bivariate linear regression (Table 2).

**Table 2 Independent variables and dependent variables in four sets of regression analysis**

| Type | Argument | Lag variable | Parameter | Equation |
|---|---|---|---|---|
| **Basic model** | $t$ | $\ln(D_{max}/D(t)-1)$ | $a=\ln(D_{max}/D_0-1)$, $k$ | Equations (2) |
| **Random sequence** | $D(t)$, $D(t)^2$ | $\Delta D(t)/\Delta t$ | $k$, $c=k/D_{max}$ | Equations (7) |
| **Regular sequence** | $D(t)$, $D(t)^2$ | $D(t+\Delta t)$ | $b=1+k\Delta t$, $c=k\Delta t/D_{max}$ | Equations (12) |
| **Continuous sequence** | $D(t)$, $D(t)^2$ | $D(t+1)$ | $b=1+k$, $c=k/D_{max}$ | Equations (16) |

## 3 Empirical analysis

### 3.1 Case of Baltimore

Two cities can be taken as examples to make case study. One is Baltimore, and the other is Shenzhen. The former represents Western cities, while the latter represents Chinese cities. The reason for choosing these two cities is because there are available sample paths of time series of fractal dimension data on urban growth in the literature. The first application case is the city of Baltimore, the most populous city in the state of Maryland of the United States of America (USA). Shen (2002) has worked out a set of fractal dimension of urban form using box-counting method. This set of fractal dimension data bear a larger time span, from 1792 to 1992. The disadvantage lies in two aspects: firstly, there are relatively few numerical points, and secondly, the time interval is uneven. However, for a case study of an algorithm, this dataset is sufficient. According to the models derived above and related principle, the data of Shen (2002) can be processed as below (Table 3). Using the processed data, we can used at least two approaches to estimate the parameter values of logistic models of fractal dimension curve. There are slight differences in parameter estimation results based on different methods (Table 4).

**Table 3 Data preparation for parameter estimation of the logistic model of fractal dimension curve of urban form of Baltimore**

| Year $n$ | Time $t$ | Fractal dimension $D$ | $D(t\text{-}1)$ | $D(t\text{-}2)^2$ | d$D(t)$/d$t$ | $D_0$ | ln($D_{max}/D(t)\text{-}1$) |
|---|---|---|---|---|---|---|---|
| 1792 | 0 | 0.6641 | 0.6641 | 0.4410 | 0.0117 | 0.6641 | 0.6638 |
| 1822 | 30 | 1.0157 | 1.0157 | 1.0316 | 0.0048 | 0.8253 | -0.0794 |
| 1851 | 59 | 1.1544 | 1.1544 | 1.3326 | 0.0019 | 0.7821 | -0.3674 |
| 1878 | 86 | 1.2059 | 1.2059 | 1.4542 | 0.0044 | 0.6714 | -0.4776 |
| 1900 | 108 | 1.3024 | 1.3024 | 1.6962 | 0.0032 | 0.6398 | -0.6928 |
| 1925 | 133 | 1.3836 | 1.3836 | 1.9143 | 0.0041 | 0.5838 | -0.8864 |
| 1938 | 146 | 1.4374 | 1.4374 | 2.0661 | 0.0105 | 0.5704 | -1.0236 |
| 1953 | 161 | 1.5953 | 1.5953 | 2.5450 | 0.0038 | 0.6866 | -1.4927 |
| 1966 | 174 | 1.6450 | 1.6450 | 2.7060 | 0.0062 | 0.6910 | -1.6726 |
| 1972 | 180 | 1.6822 | 1.6822 | 2.8298 | 0.0034 | 0.7235 | -1.8233 |
| 1982 | 190 | 1.7163 | 1.7163 | 2.9457 | 0.0005 | 0.7342 | -1.9775 |
| 1992 | 200 | 1.7211 | 1.7211 | 2.9622 | -- | 0.6856 | -2.0007 |

**Note**: The fractal dimension values in the third column come from Shen (2002). The other data are processed in this paper.

**Table 4 Parameter estimation results of the logistic model of fractal dimension curve of urban form of Baltimore**

| Parameter/Statistic | Result 1 (Approach 1) | Result 2 (Approach 2) | Result 3 (Other) |
|---|---|---|---|
| Capacity $D_{max}$ | 1.9539 | 1.9539 | 2.0000 |
| Coefficient $D_{max}/D_0\text{-}1$ | 1.8393 | 1.7152 | 1.7365 |
| Initial growth rate $k$ | 0.0131 | 0.0125 | 0.0118 |
| Initial fractal dimension $D_0$ | 0.6882 | 0.7196 | 0.7309 |
| Goodness of fit $R^2$ | 0.7142 | 0.9635 | 0.9658 |

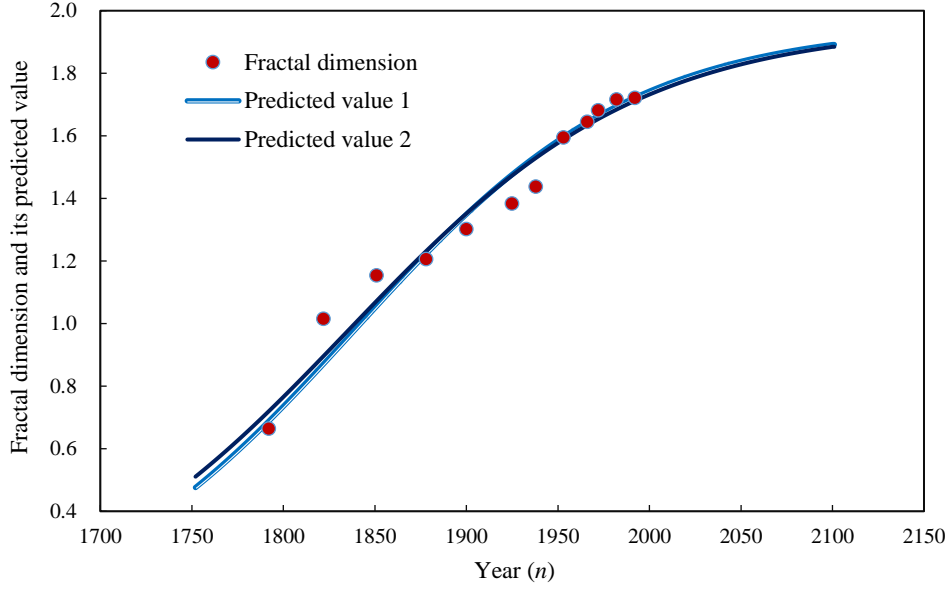

**Figure 1. The observed and predicted values of fractal dimension of urban growth of Baltimore, USA**

[**Note**: "Fractal dimension" indicates the scatter points of observed values, "predicted value 1" is the trend line based on approach 1, and "predicted value 2" is the trend line based on approach 2. The predicted curve based on the brute force search method was not displayed considering the comparability of the results. The same below.]

Taking the city of Baltimore as example, we can build the logistic model of fractal dimension curve step by step by different parameter estimation approaches. For the first approach, the procedure is as follows. Step 1: processing data. The results are shown in Table 3. Step 2: nonlinear auto-regression. Using $dD(t+1)/dt$ as dependent variable, and using $D(t)$ and $D(t)^2$ as two independent variables, letting the intercept be zero, we can make a bivariate linear regression analysis. The model is equation (7). The goodness of fit of regression analysis is about $R^2$=0.7142. This process is actually a nonlinear auto-regression. By this regressive analysis, we obtain the estimated values of $c$ and $k$. The results are $c$=0.006694 and $k$=0.01308. Step 3: parameter estimation. Using equation (8), we can estimate the value of the capacity parameter, $D_{max}$= $k/c$ =1.9539. Using equation (9) and (10), we can estimate the value of the initial fractal dimension, $D_0$=0.6882. Thus, $a$= $D_{max}/D_0$-1=1.8393 (Table 4). Step 4: model building. Substituting the estimated value of parameters into equation (1) yields a logistic growth model as follows

$$\hat{D}(t) = \frac{1.9539}{1+(1.9539/0.6882-1)e^{-0.0131t}} = \frac{1.9539}{1+1.8393e^{-0.0131t}}, \tag{21}$$

by which we can give the first set of predicted value of fractal dimension of Baltimore's urban form (Table 5).

For the second approach, the procedure is as follows. Step 1: processing data. The method and results are the same as above. Step 2: nonlinear auto-regression. The method and results are still the same as above. Step 3: parameter estimation. Give up the value of $k$. Based on the estimated value of $D_{max}$, another regressive analysis can be made by using equations (2), (3), and (4) to estimate the value of $a$ and $k$. The goodness of fit is $R^2$=0.9635. The results are ln$a$=0.5396 and $k$=0.01248. Thus $a$=exp(0.5396)=1.7152. Using equation (5), we can estimate the value of the initial fractal dimension, $D_0$=0.7196 (Table 3). Step 4: model building. Substituting the estimated value of parameters into equation (1) yields another logistic growth model as below

$$\hat{D}(t) = \frac{1.9539}{1+(1.9539/0.7196-1)e^{-0.0125t}} = \frac{1.9539}{1+1.7152e^{-0.0125t}}, \tag{22}$$

by which we can generate the second set of predicted value of fractal dimension of Baltimore's urban form (Table 4). In contrast, the brute force search method yields the following model

$$\hat{D}(t) = \frac{2}{1+(2/0.7309-1)e^{-0.0118t}} = \frac{2}{1+1.7365e^{-0.0118t}}. \tag{23}$$

The goodness of fit of regression analysis is still $R^2$=0.9658. Using the above models, equations (21) and (22), to calculate the predicted values and matching them with the observed values, the results show that the fitting effect is good (Figure 1). Based on the first approach, the value of the sum of squares of errors (SSE) is 0.0528, which is greater than the SSE value based on the second approach, 0.0457. This suggests the model of Baltimore's urban form dimension based on the second approach is better than the model based on the first approach.

**Table 5 Predicted values and corresponding errors of the logistic model of fractal dimension curve of urban form of Baltimore**

| Year $n$ | Time $t$ | Fractal dimension $D$ | Predicted value 1 | | Predicted value 2 | |
|---|---|---|---|---|---|---|
| | | | Predicted $D$ | Error $e$ | Predicted $D$ | Error $e$ |
| 1792 | 0 | 0.6641 | 0.6882 | -0.0241 | 0.7196 | -0.0555 |
| 1822 | 30 | 1.0157 | 0.8713 | 0.1444 | 0.8964 | 0.1193 |
| 1851 | 59 | 1.1544 | 1.0560 | 0.0984 | 1.0727 | 0.0817 |
| 1878 | 86 | 1.2059 | 1.2233 | -0.0174 | 1.2315 | -0.0256 |
| 1900 | 108 | 1.3024 | 1.3494 | -0.0470 | 1.3515 | -0.0491 |

| 1925 | 133 | 1.3836 | 1.4768 | -0.0932 | 1.4732 | -0.0896 |
| 1938 | 146 | 1.4374 | 1.5355 | -0.0981 | 1.5295 | -0.0921 |
| 1953 | 161 | 1.5953 | 1.5963 | -0.0010 | 1.5884 | 0.0069 |
| 1966 | 174 | 1.6450 | 1.6434 | 0.0016 | 1.6342 | 0.0108 |
| 1972 | 180 | 1.6822 | 1.6633 | 0.0189 | 1.6537 | 0.0285 |
| 1982 | 190 | 1.7163 | 1.6942 | 0.0221 | 1.6840 | 0.0323 |
| 1992 | 200 | 1.7211 | 1.7223 | -0.0012 | 1.7118 | 0.0093 |

**Note**: Based on the first approach, the value of the sum of squares of errors (SSE) is 0.0528; based on the second approach, the SSE value is 0.0457. The predicted values based on the brute force search method were not provided considering the comparability of the results. The same below.

## 3.2 Case of Shenzhen

To illustrate the applicative process of model parameter estimation methods, another example may be provided for reference. The second case is Shenzhen, an important city of southern Guangdong province, People's Republic of China, located on the east bank of the Pearl River estuary. The fractal dimension values have been figured out by Dr. Man (Man and Chen, 2020). The time span of this set of fractal dimension data is from 1986 to 2017 (Table 6). The advantage of this data lies in that the quality of remote sensing images used for calculating fractal dimensions is good. The disadvantage is that the sample path is short. The research area can be divided into two scales, one is about urbanized area (UA), and the other is around metropolitan area (MA).

**Table 6 Data preparation for parameter estimation of the logistic model of fractal dimension curve of Shenzhen's urbanized area**

| Year $n$ | Time $t$ | Fractal dimension $D$ | $D(t-1)$ | $D(t-2)^2$ | $dD(t)/dt$ | $D_0$ | $\ln(D_{max}/D(t)-1)$ |
|---|---|---|---|---|---|---|---|
| 1986 | 0 | 1.4353 | 1.4353 | 2.0600 | 0.0126 | 1.4353 | -1.1474 |
| 1989 | 3 | 1.4730 | 1.4730 | 2.1697 | 0.0448 | 1.3905 | -1.2598 |
| 1992 | 6 | 1.6073 | 1.6073 | 2.5834 | 0.0198 | 1.4728 | -1.7347 |
| 1995 | 9 | 1.6668 | 1.6668 | 2.7783 | 0.0066 | 1.4838 | -2.0067 |
| 1998 | 12 | 1.6866 | 1.6866 | 2.8447 | 0.0093 | 1.4396 | -2.1111 |
| 2001 | 15 | 1.7145 | 1.7145 | 2.9394 | 0.0246 | 1.4134 | -2.2739 |
| 2003 | 17 | 1.7636 | 1.7636 | 3.1102 | 0.0083 | 1.4799 | -2.6285 |
| 2006 | 20 | 1.7884 | 1.7884 | 3.1984 | 0.0042 | 1.4777 | -2.8595 |
| 2010 | 24 | 1.8053 | 1.8053 | 3.2593 | 0.0040 | 1.4353 | -3.0495 |
| 2013 | 27 | 1.8174 | 1.8174 | 3.3028 | 0.0068 | 1.4071 | -3.2073 |
| 2015 | 29 | 1.8309 | 1.8309 | 3.3523 | 0.0079 | 1.4259 | -3.4189 |
| 2017 | 31 | 1.8467 | 1.8467 | 3.4105 | | 1.4785 | -3.7337 |

**Note**: The fractal dimension values in the third column come from Man and Chen (2020). The other data are processed in this paper.

Let see the concrete model building results of Shenzhen city based on different parameter estimation approaches. For the fractal dimension curve within the urbanized area, the model based on the first approach, nonlinear autoregression plus formula averaging method, is as below

$$\hat{D}(t) = \frac{1.8909}{1 + (1.8909 / 1.4450 - 1)e^{-0.0131t}} = \frac{1.8909}{1 + 0.3086 e^{-0.0793t}}. \tag{24}$$

The multiple correlation coefficient square of regression analysis is about $R^2$=0.7537. Using the second approach, nonlinear autoregression plus linear regression analysis, gives a model as follows

$$\hat{D}(t) = \frac{1.8909}{1 + (1.8909 / 1.4419 - 1)e^{-0.0125t}} = \frac{1.8909}{1 + 1.3114 e^{-0.0799t}}. \tag{25}$$

The goodness of fit of the second regression analysis is about $R^2$=0.9870. In contrast, based on the brute force search method, the model is

$$\hat{D}(t) = \frac{1.8869}{1 + (1.8869 / 1.4380 - 1)e^{-0.0823t}} = \frac{1.8869}{1 + 0.3121 e^{-0.0823t}}. \tag{26}$$

The goodness of fit of regression analysis is $R^2$=0.9872. All the model parameter values and related statistics are tabulated for comparison (Table 7).

**Table 7 Parameter estimation results of the logistic model of fractal dimension curve of urban form of Shenzhen's urbanized area**

| Parameter/Statistic | Result 1 (Approach 1) | Result 2 (Approach 2) | Result 3 (Other) |
|---|---|---|---|
| Capacity $D_{max}$ | 1.8909 | 1.8909 | 1.8869 |
| Coefficient $D_{max}/D_0$-1 | 0.3086 | 0.3114 | 0.3121 |
| Initial growth rate $k$ | 0.0793 | 0.0799 | 0.0823 |
| Initial fractal dimension $D_0$ | 1.4450 | 1.4419 | 1.4380 |
| Goodness of fit $R^2$ | 0.7537 | 0.9870 | 0.9872 |

**Note**: The third set of result is based on brute force search method.

It can be seen that the modeling processes and results of Shenzhen's fractal dimension curve are similar to those of Baltimore's fractal dimension curve. Using the first and the second model, equations (24) and (25), to generate the predicted values and matching the predicted values with the observed values, the results show that the fitting effect is good, and the two predicted curves have no clear difference (Figure 2, Table 8). Based on the first approach, the value of SSE, 0.0040, is greater than the SSE value based on the second approach, 0.0038. This suggests, where urbanized area is concerned, the model of Shenzhen's urban form dimension based on the second approach is better than the model based on the first approach.

**Table 8 Predicted values and corresponding errors of the logistic model of fractal dimension curve of urban form of Shenzhen's urbanized area**

| Year $n$ | Time $t$ | Fractal dimension $D$ | Predicted value 1 | | Predicted value 2 | |
|---|---|---|---|---|---|---|
| | | | Predicted $D$ | Error $e$ | Predicted $D$ | Error $e$ |
| **1986** | 0 | 1.4353 | 1.4450 | -0.0097 | 1.4419 | -0.0066 |
| **1989** | 3 | 1.4730 | 1.5209 | -0.0479 | 1.5188 | -0.0458 |
| **1992** | 6 | 1.6073 | 1.5866 | 0.0207 | 1.5853 | 0.0220 |
| **1995** | 9 | 1.6668 | 1.6425 | 0.0243 | 1.6419 | 0.0249 |
| **1998** | 12 | 1.6866 | 1.6895 | -0.0028 | 1.6894 | -0.0027 |
| **2001** | 15 | 1.7145 | 1.7284 | -0.0140 | 1.7286 | -0.0142 |
| **2003** | 17 | 1.7636 | 1.7505 | 0.0131 | 1.7508 | 0.0127 |
| **2006** | 20 | 1.7884 | 1.7784 | 0.0100 | 1.7789 | 0.0095 |
| **2010** | 24 | 1.8053 | 1.8076 | -0.0023 | 1.8082 | -0.0029 |
| **2013** | 27 | 1.8174 | 1.8246 | -0.0073 | 1.8253 | -0.0079 |
| **2015** | 29 | 1.8309 | 1.8341 | -0.0031 | 1.8347 | -0.0037 |
| **2017** | 31 | 1.8467 | 1.8422 | 0.0046 | 1.8428 | 0.0040 |

**Note**: The fractal dimension values come from Man and Chen (2020). Based on the first approach, the value of SSE is 0.0040; based on the second approach, the SSE value is 0.0038.

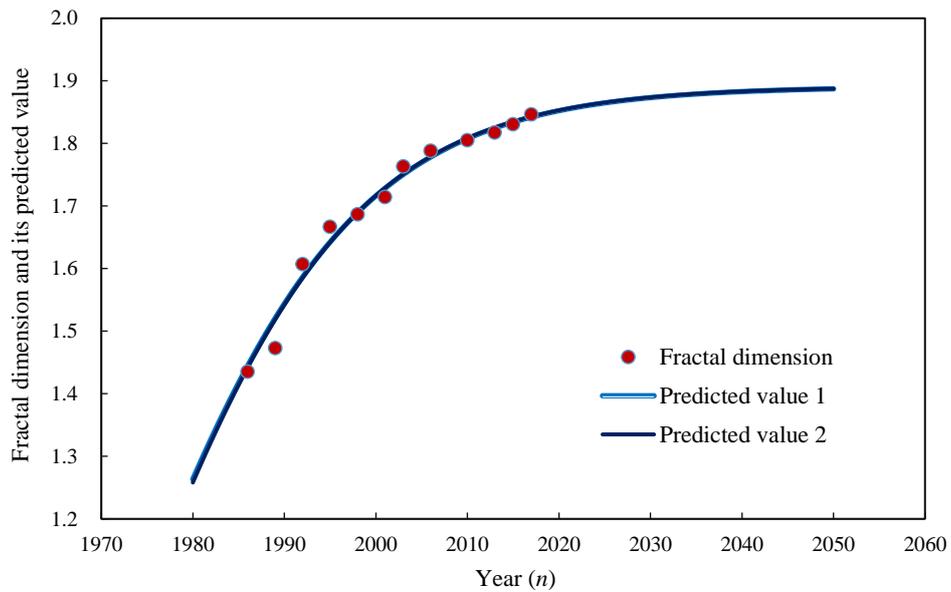

**Figure 2. The observed and predicted values of fractal dimension of urban growth of urbanized area of Shenzhen, China**

Similar methods can be used for modeling and analyzing fractal dimension curves of urban form and growth within the Shenzhen metropolitan area. In order to save space, the specific results will not be provided in detail anymore. All data and modeling results can be found in the supplementary files (File S1). The metropolitan area of a city is greater than the urbanized area of the city. Therefore, the fractal dimension value of urban form in metropolitan area is less than those in urbanized area. Generally speaking, the fractal development in urbanized areas is better than that in metropolitan area (Benguigui *et al*, 2000). For urbanized area, different approaches give similar model parameter values and predicted values (Figure 2). For metropolitan area, there are slight differences in the model parameter values and predicted values given by different methods (Figure 3). Where the logistic modeling of fractal dimension curves of Shenzhen's metropolitan area is concerned, the second approach seems to be more preferable than the first approach.

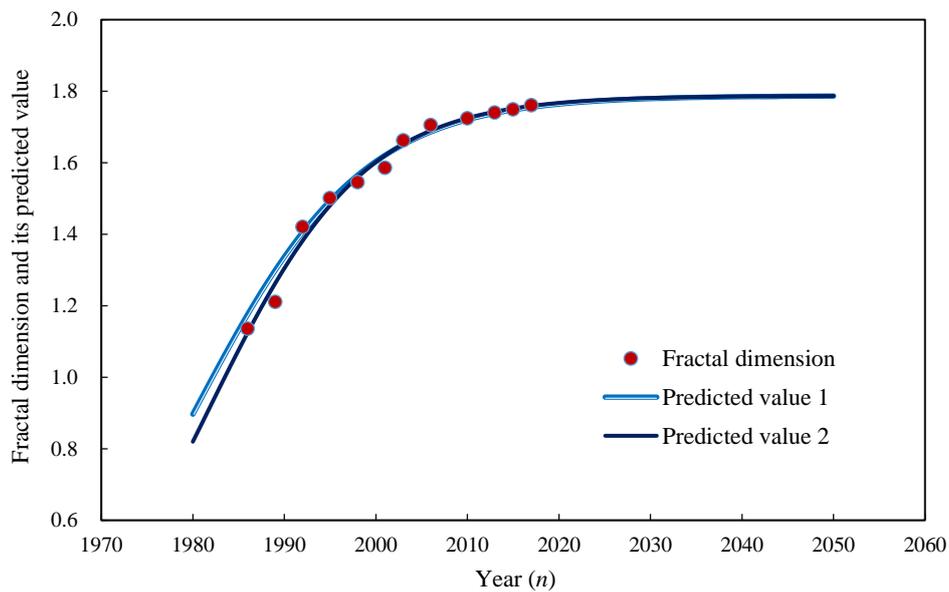

**Figure 3. The observed and predicted values of fractal dimension of urban growth of metropolitan area of Shenzhen, China**

# 4 Discussion

The results show that the nonlinear auto-regression analysis is one of effective approach for estimating the parameter values of logistic model of urban fractal dimension curves. The method is simple and suitable for three-parameter conventional logistic models. As a matter of fact, there are other methods which can be utilized to estimate the parameter value of logistic models, including curve-fitting method and brute force search method. Each method has its own advantages and disadvantages. Therefore, each method has its own scope of application. Comparing the advantages and disadvantages of different methods helps us choose appropriate methods based on specific situations (Table 9). To facilitate readers' drawing a comparison between different methods, it may be helpful to briefly explain the brute force search method here. It can be conducted by means of equation (2). If the capacity value of fractal dimension, $D_{max}$, can be determined, then a simple linear regressive analysis can be made to estimate the value of $a$ and $k$. In theory, $D_{max}$ comes between the maximum value of fractal dimension sequence, $\max(D(t))$, and the Euclidean dimension of the embedding space, $d=2$. Therefore, we can search a value between $\max(D(t))$ and $d$ so that the goodness of fit is close to the maximum value. By observing $R$-squared value while adjusting the estimated value of capacity parameter $D_{max}$ and conducting repeated experiments, the desired $D_{max}$ value can be quickly found based on a certain degree of accuracy (Chen, 2018).

**Table 9 A comparison between different methods for parameter estimation of logistic model of fractal dimension curve of urban growth**

| Method | Strength | Weakness | Scope of application |
|---|---|---|---|
| **Curve-fitting method** | There are ready-made packages available | Significant error in estimating capacity parameters | Prediction analysis using logistic model as a black box without requiring specific parameter meanings |
| **Brute force method** | This method can be applied to a logistic curve regardless of whether it is smooth or not | The operation process is cumbersome and sometimes overestimates capacity parameters | Replacement dynamic processes with significant random disturbances |
| **Nonlinear auto-regression** | For smooth logistic curves, the estimation effect of capacity | The large fluctuation in the later stage of the sequence affects the estimation of | Stable replacement dynamic processes with insignificant random |

| method | parameters is better | capacity parameters | disturbances |
|--------|---------------------|---------------------|--------------|

Logistic modeling of fractal dimension curves depends on fractal dimension measurement can calculation methods. There are many methods for calculating the fractal dimension of urban form. Among various methods for fractal dimension estimation, there are two commonly used methods: one is box-counting method (Benguigui *et al*, 2000; Jiang and Liu, 2012; Shen, 2000; Sun and Southworth, 2013), and the other is radius-area scaling method (Batty, 1991; Batty and Longley, 1994; Jiang and Zhou, 2006; White and Engelen, 1993). The box method can be generalized to grid method (Frankhauser, 1994), while in theory, the radius-area scaling method can be replaced by radius-density scaling method or radius-number scaling method (Longley *et al*, 1991; Sambrook and Voss, 2001). The box dimension represents a global fractal parameter, while the radial dimension is a local fractal parameter (Frankhauser, 1998a; Frankhauser, 1998b). The former reflect space filling, while the latter reflect spatial diffusion. In theory, under ideal conditions, the box dimension may be equal to radial dimension (Batty and Longley, 1994). In this sense, the time series of fractal dimension based on different calculation methods can be modeled with sigmoid function. However, in practice, the fractal dimension sequences obtained by means of different measurement methods reflect different effects of logistic growth modeling. For box-counting method, fractal dimension value relies on the definition of study area (measure area). For radius-area/density scaling method, the fractal dimension depends on the selection of measurement center. Using box-counting method to measure fractal dimension, we have two ways of abstracting data. One is the fixed box method (Batty and Longley, 1994; Longley *et al*, 1991; Shen, 2002), and the other is the variable box method (Benguigui *et al*, 2000; Feng and Chen, 2010). Both the two types of fractal dimension sequences can reflect urban growth, but the former can better reflect space filling of urban administrative, while the latter can better reflect space filling of built-up area. In terms of logistic modeling analysis, the fractal dimension sequence based on fixed box method performs better.

The parameter estimation method based on nonlinear auto-regression can be extended to logistic growth models in many fields. Logistic growth equation can be derived from the principle of general system theory (Bertalanffy, 1968). The logistic growth is a common phenomenon that can be seen almost everywhere. Logistic model reflects replacement dynamics process (Fisher and Pry, 1971; Hermann and Montroll, 1972; Rao *et al*, 1989). The realistic logistic process includes urbanization (Cadwallader, 1996; Karmeshu, 1988; United Nations, 2004; Zhou, 1995), urban space filling (Chen, 2012), traffic network development (Chen, 2023), technological change (Fisher and Pry, 1971), epidemic spread, and cumulative acceptance of an innovation in time (Morrill *et al*, 1988). The method developed in this article may be applicable to any growth process that can be modeled using logistic functions. The previous research similar to this work has not been seen in the literature. The nonlinear auto-regression analysis was preliminarily developed for estimating the logistic model parameters of urbanization curve (Chen, 2009). There are similarities and comparability between urbanization curve and fractal dimension curve. By analogy, the method can be developed and applied to the parameter estimation of logistic model of fractal dimension curves (Chen, 2018). Compared with previous works, the novelty of this study rests with four aspects. (1) First, a set of equations and mathematical formulae for parameter estimation of logistic models of fractal dimension curves are derived. (2) The approaches of the related parameter estimation are systematized. (3) A comparison is drawn between different approaches. (4) Two sets of case analyses are provided to illustrate the newly developed methods.

The main shortcomings of this studies lies in three aspects. Frist, if there is significant disturbance in the fractal dimension sequence, the effectiveness of this method is limited. This method developed in this paper is suitable for the fractal dimension sequences that are relatively stable and have no significant fluctuations. If the fractal dimension curve is a smooth S-shaped curve, then using this method has a good effect. If the fractal dimension curve fluctuates up and down so that the fractal dimension growth rate includes negative values, the effect of the method may be not good or even applicable. Second, lack of sample path based on regular sampling time series. The precise maps and remote sensing images suitable for calculating the fractal dimension of urban morphology are not always available. Therefore, the time for calculation of fractal dimension of a city is often irregular. In this case, I only illustrate the parameter estimation process based on random sampling sequence rather than regular sampling and continuous sampling sequences. Third, the parameter estimation methods advanced in this paper are only suitable for the ordinary logistic model. The fractal dimension increase of urban form bears squashing effect and can be described with sigmoid function. The family of general sigmoid function include the ordinary logistic function, quadratic logistic function, fractional logistic function, Boltzmann equation, and so on (Chen and Huang, 2019). Only the ordinary logistic model is take into account in this study. Further development is needed for parameter estimation methods of other sigmoid functions in future.

# 5 Conclusions

The logistic model of the fractal dimension curve of urban form has many uses. It can be used to predict urban growth peaks, urban land prospects, and urban development carrying capacity. It can be further used for dividing the stages of urban growth. It can also be used to construct logit regression models that explain the driving factors of urban growth. Such studies rely on the estimation of model parameters. If the logistic model has two parameters, that is, the capacity parameter is known, the parameter estimation process is simple. However, if the logistic model bears three parameters, namely, the capacity parameter value is unknown, the method developed in this paper can be made use. The main conclusions can be drawn as follows. First, the parameters of the general logistic model of fractal dimension curve of urban form and growth can be calculated by using the nonlinear auto-regression analysis. Based on different sampling conditions and results, the nonlinear autoregressive processes can be divided into three types, and each type contains at least two practical approaches to estimating the parameter value of the logistic model. In concrete application, different approaches can be selected according to different situations. According to the results of empirical analyses, among the two approaches, formula averaging method and regression analysis method, the most desirable approach are the second one. Second, this method can be extended to parameter estimation of logical models in other fields. If a growing process is subject to squashing effect, it can be modeled by logistic function. If a growing curve can be described with logistic function, the parameter values can be estimated by using the methods proposed in this paper. Generally speaking, urbanization curve, beta index curve of transport network, cumulative acceptance an innovation in time, and so on, take on squashing effect and can be modeled by logistic function. Therefore, the approach based on nonlinear auto-regression is applicative to the growth models of urbanization, transport network, and technology innovation diffusion.

**Acknowledgement:**

This research was sponsored by the National Natural Science Foundation of China (Grant No. 42171192). The support is gratefully acknowledged.